# Calculation of the bulk modulus of mixed ionic crystal $NH_4Cl_{1-x}Br_x$

VASSILIKI KATSIKA-TSIGOURAKOU

*Section of Solid State Physics, Department of Physics, National and Kapodistrian University of Athens, Panepistimiopolis, 157 84 Zografos, Greece*

E-mail: vkatsik@phys.uoa.gr

**Abstract**. The ammonium halides present an interesting system for study in view of their polymorphism and the possible internal rotation of the ammonium ion. The static properties of the mixed ionic crystal $NH_4Cl_{1-x}Br_x$ have been recently investigated, using three-body potential model (TDPM) by the application of Vegard's law. Here, by using a simple theoretical model, we estimate the bulk modulus of their ternary alloys $NH_4Cl_{1-x}Br_x$, in terms of the bulk modulus of the end members alone. The calculated values are comparable to those deduced from the three-body potential model (TDPM) by the application of Vegard's law.

**1.0 Introduction**

Alkali metal halides are dimorphic, crystallizing in the CsCl-type crystal structure, at low temperatures and in the NaCl-type crystal structure, at high temperatures, with the exception of $NH_4F$, which crystallizes in the ZnS-type lattice [1,2]. The ammonium halides present an interesting system for study in view of their polymorphism and the possible internal rotation of the ammonium ion [3]. $NH_4Cl$ and $NH_4Br$, at room temperature, have a simple cubic space lattice of the CsCl-type, with the tetrahedral ammonium ions oriented at random with respect to the equivalent positions in the unit cell (the hydrogen atoms pointing towards one tetrahedral set of surrounding anions in some cells and towards the other set in other cells). Notable differences exist in the properties (colour centres, ionic mobilities and defect formation, elastic anisotropy) of the solids crystallizing in the two lattices. These may be presumably of a structural origin and there is clearly a need for a better understanding of the cohesion in the salts of the CsCl-type [4]. Because of the ionic character of binding of ammonium halides, the researchers concentrate on their static and dynamical properties [4.5].

The cohesive energies of ammonium mixed halides, have been earlier studied by a number of authors, i.e. Ladd and Lee [6], Thakur and Sinha [7], Shukla et al. [8]. Very recently, by employing the three body potential model (TBPM) [9], Rawat et al. [1] also proceeded to such an investigation. The present paper is focused on that recent investigation. From X-ray structure analysis it has been observed that the mixed ionic crystals are a mixture of pure components and are truly crystalline and their lattice constants change linearly with concentration from one pure member to another. So, pseudo-experimental data for mixed compounds can be generated by applying Vegard's law to experimental values available for end point members. D. Rawat et al.[1] studied the mixed system $NH_4Cl_{1-x}Br_x$ successfully using TBPM and also calculated the thermophysical properties, viz., bulk modulus, molecular force constant, reststrahlen frequency and Debye temperature using the three body potential model. The calculated bulk modulus, from the TBPM model, for the pure end members ($NH_4Cl$ and $NH_4Br$) are best suited with experimental values, as shown in their table 4. The bulk modulus B as a function of the concentration decreases from $NH_4Cl$ to $NH_4Br$. The importance of three-body interactions in potential models, has also been emphasized by others, like Sims et al. [10] and Froyen and Cohen in the case of semiconductors [11] and more recently in the case of rare-earth monotellurides [12].

The question arises whether one can determine the values of bulk modulus of a mixed system, solely in terms of the elastic data of the end members. This paper aims to answer this question. We employ here a simple model (described below in Section 2), that has been also [13] used for the calculation of the compressibility of multiphased mixed alkali halides crystals grown by the melt method using the miscible alkali halides, i.e., NaBr and KCl, which have a simple cubic space lattice of the NaCl-type and measured in a detailed experimental study by Padma and Mahadevan [14,15]. In this paper we apply, for the first time, this model to mixed systems, which have a simple cubic space lattice of the CsCl-type and in particular to the mixed ammonium halides crystals.

**2. The $cB\Omega$ model and the compressibility of the defect volume**

Here we present a model that explains how the compressibility $\kappa(=1/B)$ of a mixed system $A_xB_{1-x}$ can be determined in terms of the compressibilities of the two end members A and B, which is of interest for the purpose of the present study. Let us call the two end members $A$ and $B$ as pure components (1) and (2), respectively and label $\upsilon_1$ the volume per "molecule" of the pure component (1) ( assumed to be the major component in the aforementioned mixed system $A_xB_{1-x}$), $\upsilon_2$ the volume per "molecule" of the pure component (2). Furthermore, let denote $V_1$ and $V_2$ the corresponding molar volumes, i.e. $V_1 = N\upsilon_1$ and $V_2 = N\upsilon_2$ (where $N$ stands for Avogadro's number) and assume that $\upsilon_1 < \upsilon_2$. Defining a "defect volume" $\upsilon_{2,1}^d$ as the increase of the volume $V_1$, if one "molecule" of type (1) is replaced by one "molecule" of type (2), it is evident that the addition of one "molecule" of type (2) to a crystal containing $N$ "molecules" of type (1) will increase its volume by $\upsilon_{2,1}^d + \upsilon_1$ (see chapter 12 of Ref. [16] as well as Ref. [17] ). Assuming that $\upsilon_{2,1}^d$ is independent of composition, the volume $V_{N+n}$ of a crystal containing $N$ "molecules" of type (1) and $n$ "molecules" of type (2) can be written as:

$$V_{N+n} = N\upsilon_1 + n(\upsilon_{2,1}^d + \upsilon_1) \quad \text{or} \quad V_{N+n} = [1 + (n/N)]V_1 + n\upsilon_{2,1}^d \tag{1}$$

The compressibility $\kappa$ of the mixed crystal can be found by differentiating eq.(1) with respect to pressure which gives:

$$\kappa V_{N+n} = [1 + (n/N)]\kappa_1 V_1 + n\kappa_{2,1}^d \upsilon_{2,1}^d \tag{2}$$

where $\kappa_{2,1}^d$ denotes the compressibility of the volume $\upsilon_{2,1}^d$, i.e., $\kappa_{2,1}^d \equiv -(1/\upsilon_{2,1}^d) \times (d\upsilon_{2,1}^d/dP)_T$. Within the approximation of the hard-spheres model, the "defect–volume" $\upsilon_{2,1}^d$ can be estimated from:

$$\upsilon^d_{2,1} = (V_2 - V_1)/N \quad \text{or} \quad \upsilon^d_{2,1} = \upsilon_2 - \upsilon_1 \tag{3}$$

Thus, since $V_{N+n}$ can be determined from eq.(1) (upon considering eq.(3) ), the compressibility $\kappa$ can be found from eq.(2) if a procedure for the estimation of $\upsilon^d_{2,1}$ will be employed. In this direction, we adopt a thermodynamical model for the formation and migration of the defects in solids described below which has been of value in various categories of solids including metals, ionic crystals, rare gas solids etc [18-23] as well as in high $T_c$ superconductors [24] and in complex ionic materials under uniaxial stress [25] that emit electric signals before fracture, in a similar fashion with the signals observed [26,27] before the occurrence of major earthquakes.

According to the latter model, the defect Gibbs energy $g^i$ is interconnected with the bulk properties of the solid through the relation $g^i = c^i B \Omega$ (usually called $cB\Omega$ model) where $B$ stands for the bulk modulus ($=1/\kappa$), $\Omega$ the mean volume per atom and $c^i$ is dimensionless quantity. (The superscript $i$ refers to the defect process under consideration, e.g. defect formation, defect migration and self-diffusion activation). By differentiating this relation in respect to pressure $P$, we find that defect volume $\upsilon^i$ $[= (dg^i/dP)_T]$. The compressibility $\kappa^{d,i}$ which, defined as $\kappa^{d,i} [\equiv -(d\ell n\upsilon^i/dP)_T]$, is given by [20]:

$$\kappa^{d,i} = (1/B) - (d^2B/dP^2)/[(dB/dP)_T - 1] \tag{4}$$

We now assume that the validity of eq. (4) holds also for the compressibility $\kappa^d_{2,1}$ involved in eq. (2), i.e.,

$$\kappa^d_{2,1} = \kappa_1 - (d^2B_1/dP^2)/[(dB_1/dP)_T - 1] \tag{5}$$

where the subscript "1" in the quantities at the right side denotes that they refer to the pure component (1). The quantities $dB_1/dP$ and $d^2B_1/dP^2$, when they are not experimentally accessible, can be estimated from the modified Born model according to [16,17]:

$$dB_1/dP = (n^B + 7)/3 \quad \text{and} \quad B_1(d^2B_1/dP^2) = -(4/9)(n^B + 3) \quad (6)$$

where $n^B$ is the usual Born exponent. This is the procedure that has been successfully applied in Ref. [13] for the multiphased mixed alkali crystals.

In the case that the $n^B$ Born exponent is not accessible, Smith and Cain [29] have shown that, there is a standard expression for determining $n^B$:

$$n^B + 1 = r_0/\rho \quad (7)$$

where $r_0$ is the nearest neighbour distance and $\rho$ is the range parameter.

## 3. Application of the model to $NH_4Cl_{1-x}Br_x$ (x=0.20, 0.40, 0.60, 0.80)

Here we use the calculated values for the lattice constants ($r_0$) and the range parameter ($\rho$), which are given in table 1 of Ref. [1] and the values for the bulk modulus ($B$), which are given in table 4 of the same Reference, for the end members $NH_4Cl$ and $NH_4Br$.

Using these values of $r_0$ and $\rho$ and applying eq. (7), we find $n^B = 9.438$ for $NH_4Cl$ and $n^B = 7.561$ for $NH_4Br$ respectively. By inserting these values into eqs. (6) we find $dB/dP = 5.479$ and $d^2B/dP^2 = -0.240$ GPa$^{-1}$ for $NH_4Cl$ and $dB/dP = 4.854$ and $d^2B/dP^2 = -0.335$ GPa$^{-1}$ for $NH_4Br$ subsequently, by inserting these values into eq. (5) we find the values of the compressibility $\kappa_{2,1}^d$ of the "defect-volume":

$$\kappa_{2,1}^d = 9.706 \times 10^{-2} \text{ GPa}^{-1}, \text{ for } NH_4Cl \text{ and}$$

$$\kappa_{2,1}^d = 15.835 \times 10^{-2} \text{ GPa}^{-1}, \text{ for } NH_4Br.$$

For x=0.20, the end member (pure) crystal with the higher composition is NH$_4$Cl (component (1) ). By considering the $\upsilon_1$ and $\upsilon_2$ values of NH$_4$Cl and NH$_4$Br respectively, we find $\upsilon_{2,1}^d = 5.984 \times 10^{-24} cm^3$ from Eq. (3) and $V_{N+n} = 48.071 \times 10^{-24} cm$ from Eq. (1). By inserting the aforementioned values into Eq. (2) we find $\kappa = 4.514 \times 10^{-2}$ GPa$^{-1}$ and therefore $B = 22.15$ GPa (see table 1). This practically coincides with the value $B = 22$ GPa, reported in table 4 of Ref. [1].

For x=0.40 and considering that NH$_4$Cl is the component (1), following the same procedure as previously, we find $V_{N+n} = 66.089 \times 10^{-24} cm$ and therefrom $B = 21.41$ GPa (table 1). This slightly exceeds the $B$ value, reported in Ref. [1].

For x=0.60, the end member (pure) crystal with the higher composition is NH$_4$Br (component (1) ). Following the above procedure and the corresponding values for NH$_4$Br, as component (1), we find $V_{N+n} = 68.084 \times 10^{-24} cm$ and finally $B = 15.07$ GPa. Since this value seems to deviate markedly from the value $B = 18$ GPa, reported in Ref. [1], we repeated our calculation by considering NH$_4$Cl as component (1). In this case, we find $B = 20.75$ GPa (written in parenthesis in table 1), which again differs markedly from the value of Ref. [1].

For x=0.80 and considering that NH$_4$Br is the component (1), following the same procedure as above, we find $V_{N+n} = 52.559 \times 10^{-24} cm$ and therefrom $B = 14.50$ GPa.

## 4. Conclusion

In table 1 we present the values of the nearest neighbour distance (lattice constant, $r_0$), the range parameter ($\rho$), for NH$_4$Cl and NH$_4$Br, which are given in table 1 of Ref. [1], the bulk modulus $B$, for $NH_4Cl$ and $NH_4Br$ and the mixed system $NH_4Cl_{1-x}Br_x$, as they are presented in table 4, obtained by three-body potential model. Here, the bulk modulus for the mixed system $NH_4Cl_{1-x}Br_x$, has been estimated by using a procedure based on a simple thermodynamical model (the so called $cB\Omega$-model). This procedure leads to the bulk modulus values, which are more or less comparable with those obtained from the three-body potential model employed in Ref. [1]. Only one marked deviation for the composition $NH_4Cl_{0.40}Br_{0.60}$ has been noticed, the origin of which however is not yet clearly understood.

Concerning the above agreement between the present results and those of Ref. [1], two comments are in order.

First, the Born exponent was calculated through eq. (7). This equation in reality does the following: when the repulsive interaction energy $W_R$ is modelled either as a power law or as given by exponential form, we assume that in both procedures the stiffness of the interaction (which in Ref. [38] is defined as $\pi \equiv -r^2 W_R'' / r W_R'$ where the primes denote differentiation with respect to crystal distance $r$ and $W_R$ includes all terms in the lattice energy except the Coulomb energy) is the same. In other words, the $n^B$ value obeying eq. (7) adjusts the stiffness of the interaction to remain unaltered upon using either power law or exponential form. In addition, as already commented on in Ref. [38], note that the isothermal bulk modulus and the nearest neighbour distance are the important pieces of experimental information that go into the empirical Born model determination of $\pi$. Thus, the key point here is to consider reliable values of these two pieces of experimental information (which is the case since they are taken from Ref. [1] that agree with those obtained experimentally).

Second, the values of $r_o$ and $\rho$ we used for the application of eq. (7) come from the three body potential model on which the computations of Ref. [1] were based. In other words, in our procedure here we employed certain model parameters derived from the three body potential model.

Thus, considering the aforementioned two comments one would wonder why the results of our procedure were found to be in good agreement with those of Ref. [1]. This could be understood in the following context when considering for the sake of simplicity that the component (1) is the dominant component in the mixed crystal. Then, the key point to calculate correctly the quantity $\kappa/\kappa_1$ from eq. (2) is to determine the ratio $\kappa^d_{2,1}/\kappa_1$, which is deduced from eq. (5). This equation reveals the interesting property that the ratio $\kappa^d_{2,1}/\kappa_1$ is equal to

$$\kappa^d_{2,1}/\kappa_1 = 1 - B_1 \frac{(d^2 B_1/dP^2)}{(dB_1/dP)_T - 1} \quad (8)$$

which (irrespective of the mixed crystal concentration) is solely governed by the elastic properties of the pure component (1). These elastic properties have been successfully determined in Ref. [1], since the calculated values were found to be in agreement with the experimental data for the pure components. Eq. (7) adjusts the $n^B$ value to the $r_o$ and $\rho$ values (taken from Ref. [1], thus reproducing reliably the elastic data of the pure component (1)) and therefrom we can approximate the quantities $dB_1/dP$ and $d^2B_1/dP^2$ given by eqs. (6) which are inserted into the right hand side of eq. (8).

In conclusion, by using a simple thermodynamical model, here we estimated two values for the bulk modulus of the mixed ionic crystal $NH_4Cl_{1-x}Br_x$ considering both end members ($NH_4Cl$ and $NH_4Br$) as the dominant component (1). These values agree well with those deduced from the recent results of Ref. [1].

**Table 1.** The values of lattice constant, ($r_0$), range parameter ($\rho$) and the bulk modulus ($B$), from Ref. [1]. In addition, the last column reports the values of bulk modulus ($B$) for the mixed system $NH_4Cl_{1-x}Br_x$ as estimated with the procedure described in the text

| composition | $r_0(\overset{o}{A})^{(1)}$ | $\rho(\overset{o}{A})^{(1)}$ | $B(GPa)^{(2)}$ | $B(GPa)^{(3)}$ |
|---|---|---|---|---|
| $NH_4Cl$ | 3.34 | 0.32 | 24 | |
| $NH_4Cl_{0.8}Br_{0.2}$ | | | 22 | 22.15 |
| $NH_4Cl_{0.6}Br_{0.4}$ | | | 20 | 21.41 |
| $NH_4Cl_{0.4}Br_{0.6}$ | | | 18 | 15.07 |
| | | | | (20.75) |
| $NH_4Cl_{0.2}Br_{0.8}$ | | | 15 | 14.50 |
| $NH_4Br$ | 3.51 | 0.41 | 14 | |

[1] Literature values, which are given in table 1 of Ref. [1]

[2] Literature values, which are given in table 4 of Ref. [1]

[3] Calculated from eq. (2), by inserting $\kappa^d$ from eq. (5)